\documentclass[aps,prc,groupedaddress,showpacs,preprint]{revtex4}
\def\half{{\textstyle{1\over 2}}}

\def\fourth{{\textstyle{1\over 4}}}

\def\M{{\cal M}}

\def\R{{\cal  R}}
\def\W{{\cal  W}}

\newcommand{\be}{\begin{equation}}
\newcommand{\ee}{\end{equation}}
\newcommand{\bea}{\begin{eqnarray}}
\newcommand{\eea}{\end{eqnarray}}
\newcommand{\bra}[1]{\langle {#1} |}                        
\newcommand{\ket}[1]{| {#1} \rangle}
\usepackage{epsfig}
\usepackage{amsmath}
\usepackage{graphics}
\usepackage[dvips]{color}

\begin{document}

\title{Charge Form Factors of Quark-Model Pions II}

\author{F. Coester\footnote{corresponding author}}
\email[]{coester@anl.gov}
\affiliation{Physics Division, Argonne National Laboratory
Argonne, IL 60439, USA}

\author{W.N. Polyzou}
\email[]{polyzou@uiowa.edu}
\affiliation{Department of Physics and Astronomy University of Iowa,
Iowa City, IA 52242 USA}


\begin{abstract}

Experimental data of the pion charge form factor are well represented
by Poincar\'e invariant constituent-quark phenomenology depending on
two parameters, a confinement scale and an effective
constituent--quark mass.  Pion states are represented by
eigenfunctions of mass and spin operators and of the light-front
momenta.  An effective current density is generated by the dynamics
from a null-plane impulse current density. A simple shape of the wave
function depending only on the confinement scale is sufficient.
\end {abstract}
\pacs{12.39.-x , relativistic quark model, pion form factor, null-plane 
kinematics}

\maketitle
In a previous letter \cite{CCP} we demonstrated that simple
constituent-quark models of the pion yielded charge form factors in
agreement with data for both low $Q^2$ \cite{AMEND} and $Q^2
>1\rm{GeV}^2$.  Recently new measurements \cite{VOLMER} provided more
precise data for $Q^2<2$ GeV$^2$ and new data at higher values of
$Q^2$ are expected.  The purpose of this paper is to show that simple
relativistic constituent-quark models are consistent with all existing
data and with a relatively narrow range of form factor values for 
larger momentum transfers.

The unitary Poincar\'e representations of confined quark states are
specified by mass and spin operators together with the choice of a
kinematic subgroup \cite{KP,FC}.  
The Poincar\'e covariant, conserved, effective
current operator $I^\mu(x)$,
\bea
U(\Lambda ,a) I^{\mu} (x) U^{\dagger} (\Lambda ,a) 
&= &I^{\nu} ( \Lambda x +a ) \Lambda_{\nu}{}^{\mu} \; ,\cr 
\imath [P_\mu, I^\mu(x)]&=&\partial_\mu  I^\mu(x)=0 \; ,
\eea
is generated by the dynamics from an input that is covariant under the
kinematic subgroup. The results of \cite{CCP} were based on the use of
null-plane kinematics for the generation of an effective current
density.  With that choice the kinematic subgroup leaves the
null-plane $n\cdot x$, with $n^2=0$, invariant. With a convenient
choice of the axes the components of $n$ are given by
$n=\{1,0,0,1\}$. It is an important feature of null-plane kinematics
that the charge form factor does not depend on the pion mass.  This
feature is essential for the empirical success of constituent-quark
phenomenology applied to pion form factors.  With point-form
kinematics \cite{ALLEN} form factors obtain as functions of
$\eta:=Q^2/4m_\pi^2$, which is large for moderate values of $Q^2$.
Thus, with simple wave functions, form factors are much too small for
a realistic representation \cite{BERTRAND}. There is no intent to
approximate features of quantum field theory in the construction of
such quark models. In particular the representations of
constituent-quark states are not meant to approximate Fock-space
amplitudes and/or satisfy features of perturbative QCD \cite{BRODSKY}.
Meson mass operators of constituent-quark models may be defined by
simple spectral representations
\be
\bra{\mu,\bar \mu,\xi,\vec k_\perp}\M\ket{\vec k_\perp',\xi',\bar \mu',,\mu'}
=\sum_{n,j,\mu} \phi_{n,j,\sigma}(\mu,\bar\mu,\xi, \vec k_\perp)M_{n,j}
\phi_{n,j,\sigma}(\mu,\bar\mu,\xi, \vec k_\perp)^*.
\ee
Neither the wave function representing the pion state nor the explicit
representation, $\bra{\mu',\bar \mu',\xi',\vec
k_\perp'}I^\nu(0)\ket{\vec k_\perp,\xi,\bar \mu,\mu}$, of the current
density are observable. The observable form factor is invariant under
simultaneous unitary transformations, which may preserve the kinematic
subgroup \cite{KP,FC}.

\begin{figure}
\begin{center}
\rotatebox{00}{\resizebox{4.5in}{!}{
\includegraphics{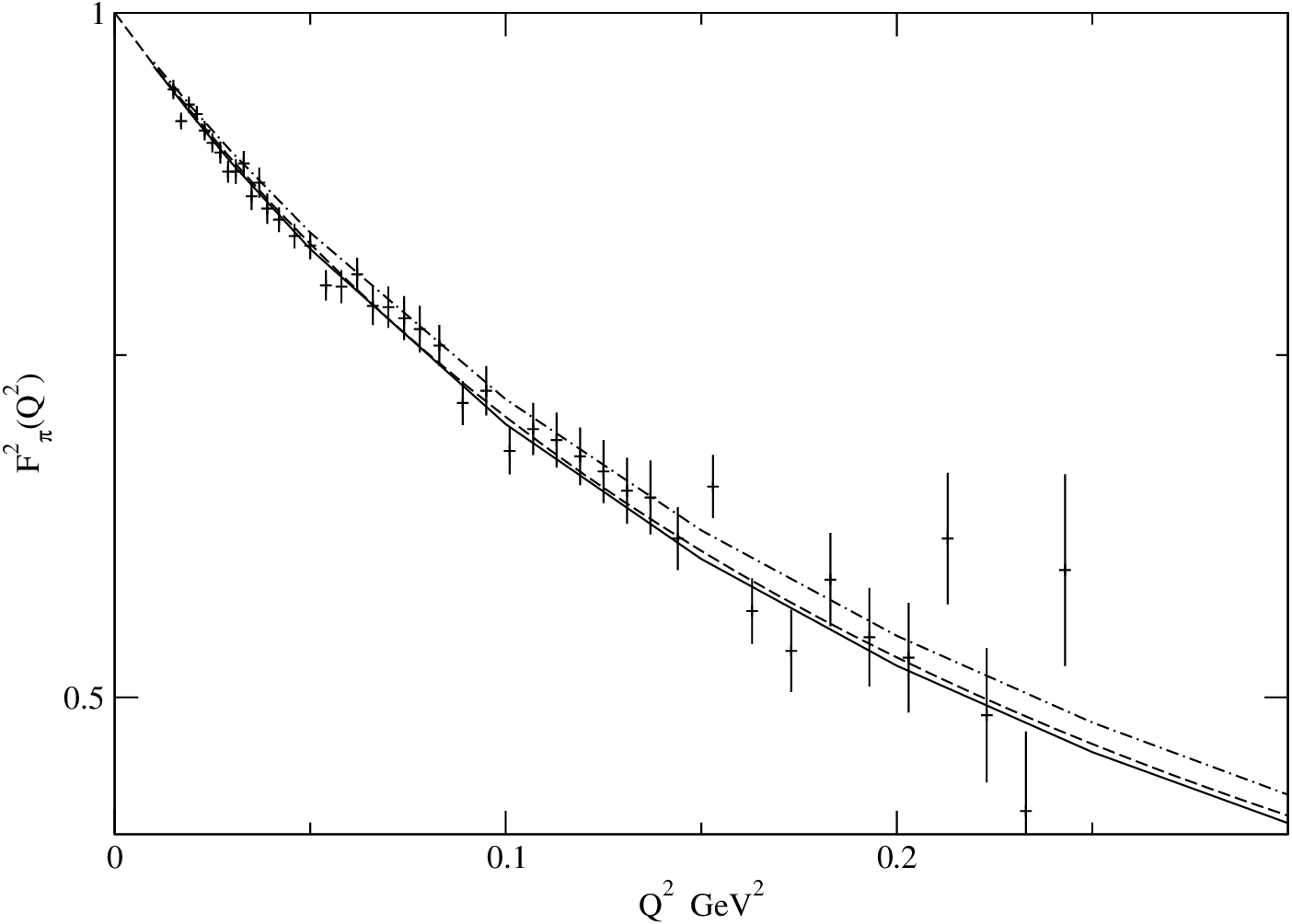}}
}
\end{center}
\caption{Wave function dependence of pion form factors at low $Q^2$
}
\end{figure}

As in ref. \cite{CCP} the models are specified by  
input currents
with the representation
\be
\bra{\xi' , k_\perp', \mu',\bar \mu'} n\cdot I(0)
\ket{\xi,  k_\perp ,\mu ,\bar \mu}
= 
 \delta_{\mu',\mu}  \delta_{\bar \mu',\bar \mu} \delta (\xi'-\xi)
 \delta( k_\perp'-  k_\perp- (1-\xi) Q_\perp) \; ,
\ee
and a representation of the pion state by wave functions
$\phi(\xi, k_\perp,\mu,\bar \mu)$ which is proportional to a
radial wave function $u(k^2)$ and Melosh rotation matrices
\be
\phi(\xi, k_\perp,\mu,\bar \mu):=\sum_{\mu',\bar\mu'}
\bra{\mu} \R^\dagger(\xi,  k_\perp)\ket{\mu'}
\bra{\bar \mu} \R^\dagger(1-\xi, - k_\perp)\ket{\bar\mu'}
(\half,\half, \mu',\bar \mu'|0,0) u(k^2)\; .
\ee
The argument $k^2$ is related to the null-plane momenta by
\be
k^2+m_q^2={ k_\perp^2+m^2\over 4 \xi(1-\xi)}\; .
\ee
The  pion charge form factor is a functional of the radial wave function, 
$u(k^2)$,
\be
F_\pi(Q^2) = {1\over 16 \pi}\int_0^1 d\xi 
\int {d^2 \bar k_\perp\over \xi(1-\xi)} \W(\xi,\bar k_\perp)
u({k'}^2)u(k^2)
\ee
with
\be
\bar k_\perp = k_\perp-\half (1-\xi)Q_\perp =k_\perp'+\half (1-\xi)Q_\perp
\ee
and
\be
{\W}:= \sqrt{{\xi(1-\xi)\over \sqrt{(m_q^2+k_\perp^2)(m_q^2+{k_\perp'}^2)}}}
{m_q^2+\bar k_\perp^2-\fourth(1-\xi)^2 Q^2\over \xi(1-\xi)}\; .
\ee

\begin{figure} 
\begin{center}

\rotatebox{00}{\resizebox{4.5in}{!}{
\includegraphics{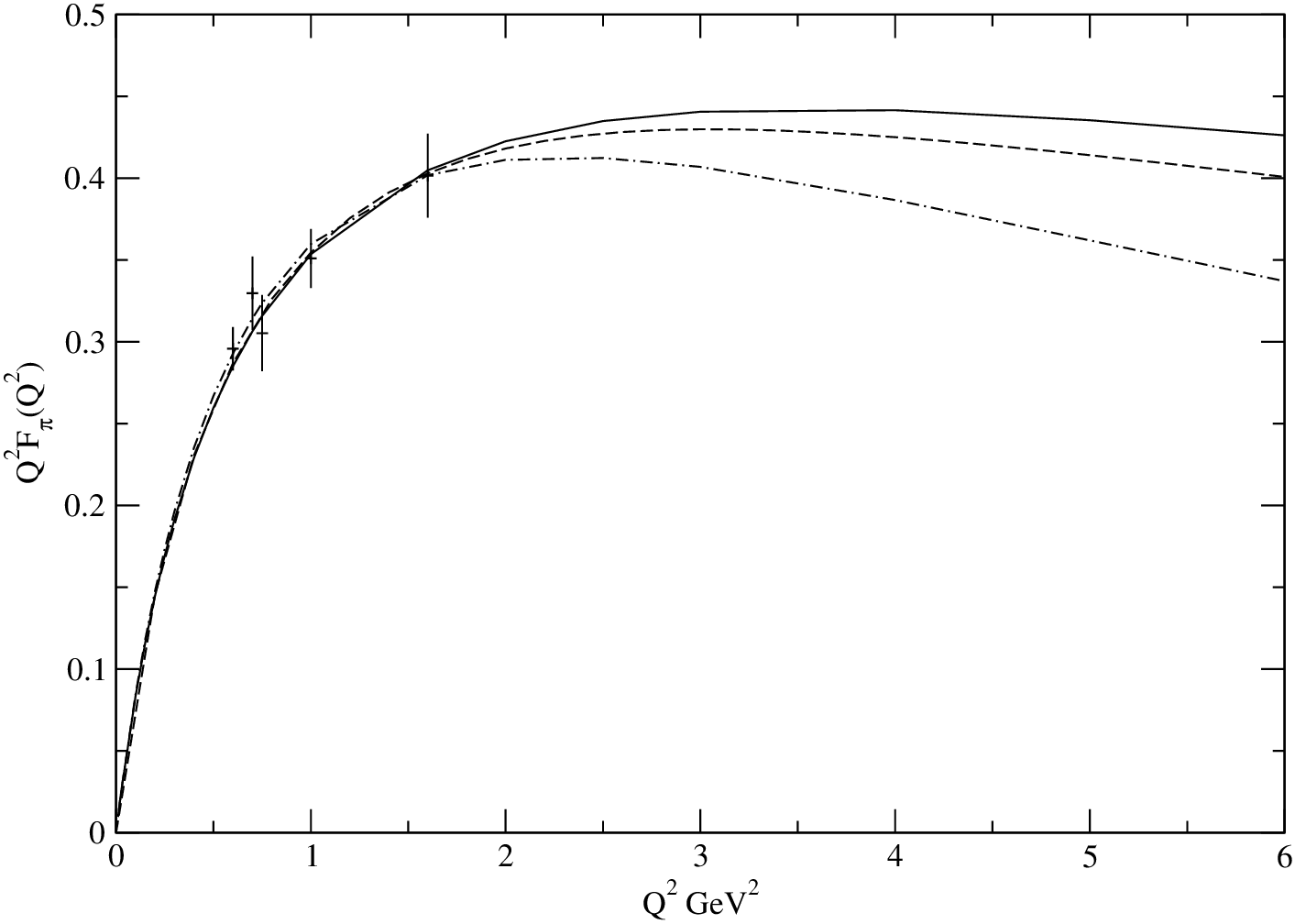}}
}
\caption{Wave function dependence of pion form factors at high $Q^2$}
\end{center}
\end{figure}

In \cite{CCP}  a Gaussian shape 
\be
u(k^2)= \sqrt{{4\over \sqrt{\pi}b^3}}\exp(-k^2/2b^2) 
\label{GAUSS}
\ee
was used for numerical convenience.  We expect that
a  rational  shape 
\be
u(k^2)= \sqrt{{32\over \pi b^3}}\left(1\over 1+k^2/b^2\right)^2
\label{FRAT}
\ee
may specify a  better model. A wave function of essentially the same shape
can also be generated by the equation \cite {POLY}
\be
\left( 2 \sqrt{m_q^2 + k^2} + \alpha r - {\beta \over r} +
\gamma\, \vec{s}_q \cdot \vec{s}_{\bar q}\,  e^{-\lambda^2 r^2}-m_0\right) u(r)=0
\label{MASS}
\ee
with the parameters adjusted for that purpose. Conventional QCD motivated mass
 operators\cite{ISGUR},
designed to fit meson spectra,  produce wave functions that require substantial
modification of the current \cite{CALDAR} .

			    \begin{figure}
\begin{center}
\rotatebox{0}{\resizebox{4.5in}{!}{
\includegraphics{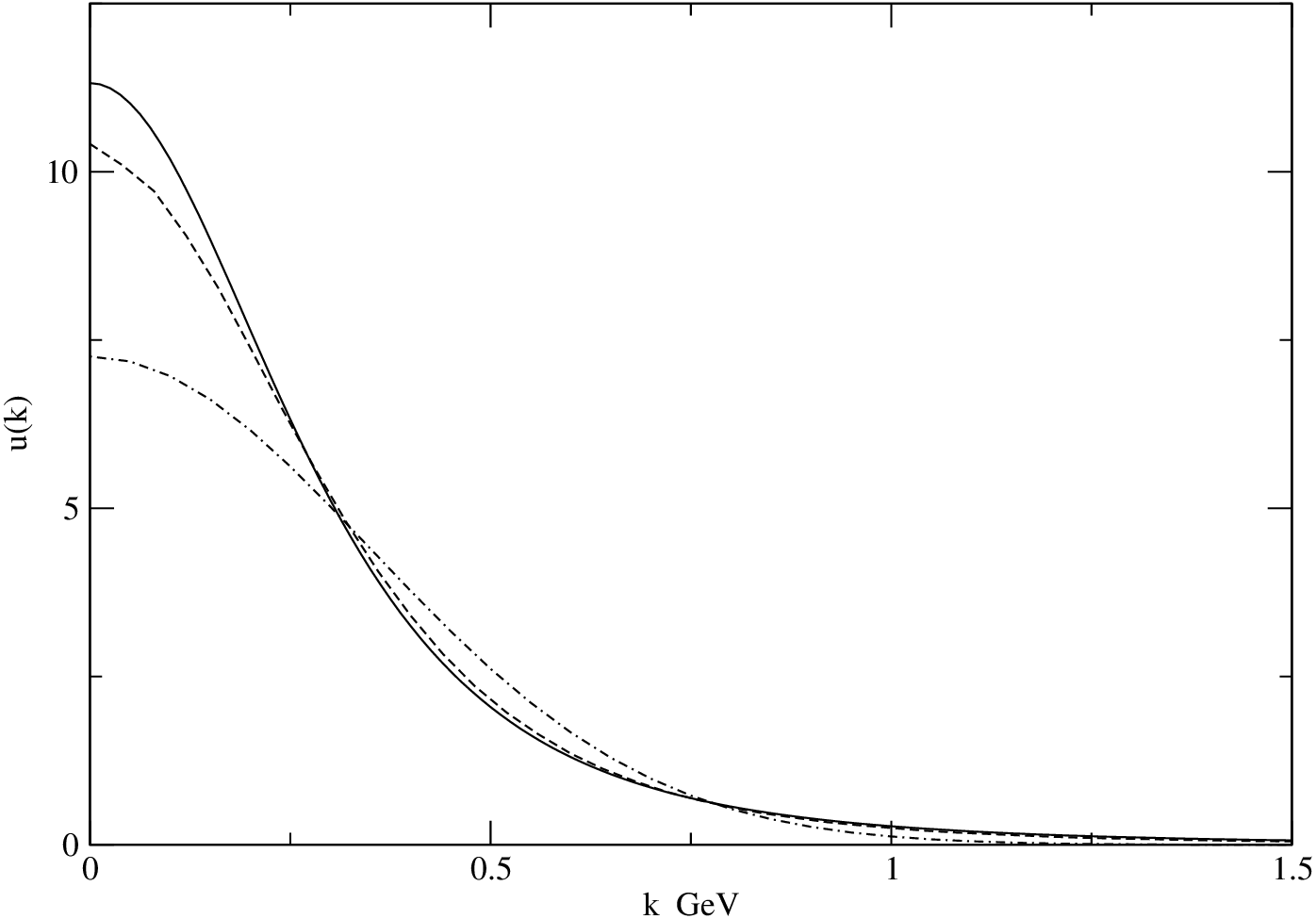}}
}
\end{center}
\caption{Shapes of the three radial wave functions}
\end{figure}

The form factors shown in Figs. 1 and 2 are calculated assuming
$m_q=.23$ GeV, $b=.35$ GeV and $.43$ GeV for the wave functions
eq. (\ref{GAUSS}) (dash-dot line) and eq. (\ref{FRAT}) (solid line)
respectively. The dash line is obtained using the ground-state
solution of eq. (\ref{MASS}) with $\alpha=.1$ GeV$^2$, $\beta=.4$, 
$\gamma=.229$ GeV and $\lambda=.3$ GeV.  
The shapes of the three wave functions are compared
in Fig.~3.  All three parameterizations are in agreement with existing
data. We expect that for larger values of $Q^2$ the form factors
obtained with the Gaussian wave function (\ref{GAUSS}) will be in
disagreement with future measurements.  The results are in agreement
with the QCD approximations of Maris and Tandy \cite{MARIS} for
$Q^2< 2 $GeV$^2$,  and with their expectations for larger values of
$Q^2$.

For the pion decay constant, \cite{CALDAR,JAUS}
\be
f_\pi= {\sqrt{3} g_A^q\over 8\pi^2}\int_0^1 d\xi 
\int {d^2 k_\perp\over \xi(1-\xi)}
{m_q\over \sqrt{M_0}} u(k^2)\; , \qquad 
M_0^2:={m_q^2+k_\perp^2\over \xi(1-\xi)}.
\ee
The three wave functions yield the  values $92.5$ MeV, $101.5$ MeV
 and $101.8$ MeV   with $g_A^q=1$.

\begin{acknowledgments}
Research supported in part by 
the U.S. Department 
of Energy, Office of Nuclear Physics  contracts  W-31-109-ENG-38 and 
DE-FG02-86ER40286 .
\end{acknowledgments}


\begin{thebibliography}{99}
\bibitem{CCP} P. L. Chung, F. Coester and W. N. Polyzou, 
Phys. Lett. B {\bf 205}, 545 (1988)
\bibitem{AMEND} S. R. Amendolia et al., Nuc. Phys. B  {\bf 277}, 168 (1986)
\bibitem{VOLMER} J. Volmer et al., Phys. Rev. Lett. {\bf 86} , 1713  (2001)
\bibitem{KP} B. D. Keister and W. N. Polyzou, Adv. Nuc. Phys. 
{\bf 20}, 225 (1991)
\bibitem{FC} F. Coester, Prog. Part. Nucl. Phys. {\bf 29}, 1 (1992)
\bibitem{ALLEN}T. W. Allen and W. H. Klink,  Phys. Rev. C {\bf 58} , 
3670 (1998)
\bibitem{BERTRAND} A. Amghar, B. Desplanques and L.Theussl, 
Phys. Lett. B {\bf 574} , 201 (2003)
\bibitem{BRODSKY} S. J. Brodsky  et al., hep-ph/0311218 
\bibitem{POLY}B. D. Keister and W. N. Polyzou, J. Comp. Phys. {\bf 134} 
231 (1997)
\bibitem{ISGUR}S. Godfrey and N. Isgur, Phys. Rev. D {\bf 32}, 189 (1985)
\bibitem{CALDAR}F. Cardarelli et al., Phys. Lett. B {\bf 332}, 1 (1994); 
{\bf 357}, 267 (1995)
\bibitem{MARIS}  P. Maris and P. C. Tandy, Phys. Rev. C {\bf 62} , 055204 (2001)
\bibitem{JAUS} W. Jaus, Phys. Rev. D {\bf 44}, 2851 (1991)
\end{thebibliography}
\end{document}